\documentstyle[12pt,epsfig]{article}

\newcommand{\be}{\begin{equation}}
\newcommand{\ee}{\end{equation}}
\newcommand{\bdis}{\begin{displaymath}}
\newcommand{\edis}{\end{displaymath}}

\newcommand{\la}{\left\langle}
\newcommand{\ra}{\right\rangle}

\newcommand{\unp}{u_n^+}
\newcommand{\unm}{u_n^-}

\begin{document}
\pagestyle{empty}
\date{}

\title{Helicity advection  in Turbulent  Models}

\author{L. Biferale$^{1,2}$, D. Pierotti$^{3}$ and F. Toschi$^{2,4}$\\
$^{1}$  Dip. di Fisica, Universit\`{a} di Tor Vergata,
 Via della Ricerca Scientifica 1,\\
I-00133, Roma, Italy.\\
$^{2}$ Istituto Nazionale di Fisica della Materia, unit\`a di Tor Vergata.\\
$^{3}$ Dept. of Chemical Physics, Weizmann Institute of Science,\\
76100, Rehovot, Israel.\\
$^{4}$ Dip. di Fisica, Universit\`{a} di Pisa, Piazza Torricelli 2,\\
I-56126, Pisa, Italy.}

\maketitle

\begin{abstract}
Helicity transfer in a shell model of turbulence is investigated.
 In particular, we study  the scaling behavior
of helicity transfer in a dynamical model of turbulence
lacking inversion symmetry. We present some phenomenological
and numerical  support to  the idea that 
Helicity becomes -at scale
small enough- a  passively-advected
quantity. 
\end{abstract}
  
\begin{flushright}
``${o}{\upsilon}{\tau}{o}{\iota} \,\,
{\sigma}{\upsilon}{\nu}{\epsilon}{\chi}{\theta}{\epsilon}{\iota}{\nu},\,\,
\alpha\lambda\lambda\alpha \,\,
\sigma\upsilon\mu\phi\iota\lambda\epsilon\iota\nu \,\,
\epsilon\phi\upsilon\nu$''\\
ANTI$\Gamma$ONH \\
``non sono nata per condividere l'odio, ma l'amore''\\
\end{flushright}
\section{Introduction}
One of the most intriguing problems in three dimensional 
fully developed turbulence (FDT) 
is related to the 
appearance of anomalous scaling laws at high Reynolds numbers, i.e. 
in the limit when Navier-Stokes dynamics is dominated by the non-linear
interactions.

The celebrated 1941 Kolmogorov theory (K41) was able to capture the
main 
phenomenological ideas by performing dimensional analysis based
on the energy transfer mechanism. 
Kolmogorov   postulated that the energy cascade should
follow a  self-similar and homogeneous process entirely dependent
on the energy transfer rate, $\epsilon$. This idea, plus the assumption of 
local isotropy and universality at small scales, 
 led  to a precise prediction on the statistical
properties of the increments of turbulent velocity fields:
$\delta v (r) \sim |v(x+r)-v(x)| \sim (r \cdot \epsilon(r))^{1/3}$.  From this the
scaling of moments of $\delta v (r)$, the structure functions,
can be determined in terms of the statistics of $\epsilon(r)$, i.e.
\be
S_p(r) \equiv \langle \left( \delta v(r)\right)^p\rangle=C_p 
\langle \left(\epsilon(r)\right)^{p/3} \rangle r^{p/3} ,
\label{eq.1}
\ee
where $C_p$ are constants and the scale $r$ is supposed to be in 
the inertial range, i.e. much smaller than the integral scale 
and much larger than the viscous dissipation cutoff.
If $S_p(r)\sim r^{\zeta(p)}$ and
 $\langle \epsilon^p(r)\rangle\sim r^{\tau(p)}$
then
\be
\zeta(p)=p/3+\tau(p/3).
\label{eq.2}
\ee

In the K41 the $\epsilon(r)$ statistic   is assumed to  be 
$r$-independent, that is $\tau(p)=0$, implying
$\zeta(p)={p\over3}, \forall p$.
On the other hand, there are many experimental and numerical
 \cite{MS,BCTBS} 
results telling  us that K41 scenario for homogeneous
and isotropic turbulence is quantitatively wrong. Strong intermittent
bursts in the energy transfer have been observed and non trivial $\tau(p)$
set of exponents measured. 

Many different authors have focused their attention on the possible 
r\^ole played by helicity, the second global invariant of 3D Navier-Stokes 
eqs. \cite{kraic,lev,waleffe,bk},  for determining leading or sub-leading 
 scaling properties of correlation functions in the inertial range.

Recently \cite{procaccia2,russo}, 
an exact scaling equation for the third order
 velocity correlations entering in the  helicity flux definition 
has been derived under two hypothesis: (i) there exists
 a non-vanishing helicity flux,
(ii) the flux becomes Reynolds independent in the limit of FDT. 
This relation 
 predicts a $r^2$-scaling for the  particular  third order  velocity 
correlation entering in the definition of helicity flux, at difference
from the celebrated linear behavior in $r$ showed by the third order
velocity correlations entering in the definition of energy flux. 
 
This simple fact tells us that different velocity correlation with the
same physical dimension but with different tensorial structure 
may show different {\em leading}  scaling properties. 

Moreover, even if overwhelming evidences indicate that the main 
physics is driven by the energy transfer, there can  be 
some sub-leading new intermittent statistics  hidden in the
helicity flux properties. \\

Homogeneous and isotropic turbulence has, by definition, always a vanishing
mean helical flux. Nevertheless, both fluctuations about the zero-mean, in
isotropic cases, and/or net non-zero fluxes, in cases where inversion
symmetry is explicitly
broken, can be of some interest for the understanding of fully developed 
turbulence. 

In this letter, we  analyze  
 the helical transfer mechanism in dynamical models
of turbulence \cite{leo,bk,bbkt},
built such as to explicitly  consider helicity conservation
in the inviscid limit. 

In a previous publication \cite{bpt} we found
   the first strong numerical evidence that a 
 Reynolds-independent helicity flux is present in cases where the forcing 
 mechanism explicitly
breaks inversion symmetry.  In this paper we present a  phenomenological
argument which supports the idea that helicity behaves as a quantity
passively transferred toward small scales by the mean energy flux,
in agreement with both numerical findings in the true Navier-Stokes eqs.
\cite{orszag} and with our previous numerical evidences about the strong
intermittent properties of the helicity flux \cite{bpt}. 

In the following, we briefly  summarize the main  motivation
behind the introduction of Shell Models for turbulence.We summarize
the main results obtained in \cite{bpt} and we   present the 
new argument about the passive character of helicity fluctuations. 

\section{Shell Models}
Shell models have demonstrated to be very useful for the understanding
of many properties connected to turbulent flows \cite{G}-\cite{ls}.
The most popular shell model, the Gledzer-Ohkitani-Yamada 
(GOY) model (\cite{G}-\cite{ls}), has been shown to predict
scaling properties for $\zeta(p)$ (for a suitable  choice of the free
parameters) similar to what is found experimentally.  

The GOY model can be seen as a severe truncation of the 
Navier-Stokes equations: 
it retains only one complex mode $u_n$ as a representative 
of all Fourier modes in 
the shell of wave numbers $k$ between $k_n=k_02^n$ 
and $k_{n+1}$.

It has been pointed out that GOY model
conserves in the inviscid, unforced limit 
two quadratic quantities. The first quantity is 
the {\it{energy}}, while the second is 
the equivalent of {\it{helicity}} in 3D turbulence 
\cite{BKLMW}.  In two recent works \cite{bk,bbkt} the GOY
model has been generalized in terms of shell variable, $u_n^+, u_n^-$,
transporting positive and negative helicity, respectively.  It is easy to 
realize that only 4 independent classes of models  can be 
derived such  as to preserve the same helical structure of Navier-Stokes 
equations \cite{waleffe}.
 All  these models have 
at least one inviscid invariants non-positive defined which is
similar to the 3D Navier-Stokes helicity.  In the following, we will
focus on the intermittent properties of one of them which has already been 
extensively investigated (see  \cite{bbkt,bbt}  for more
details). 
The time evolution for positive-helicity shells reads \cite{bbkt}:
\be
\dot{u}^+_n=i k_n \left( A_n\left[u,u\right] \right)^*
 -\nu k^2_n u^+_n +\delta_{n,n_0}f^+,
\label{eq:shells}
\ee
with the equivalent eqs, but with all helicity signs reversed, for  
$\dot{u}^-$.
In (\ref{eq:shells})
$\nu$ is the molecular viscosity, $f^+, f^-$ are the large scale forcing and  
$A[u,u]$  refers to the non-linear terms of the model. Namely:
\begin{equation}
A_n[u,u] \equiv u^{-}_{n+2} u^{+}_{n+1}+b_3
u^{-}_{n+1} u^{+}_{n-1}
+c_3 u^{-}_{n-1} u^{-}_{n-2}.
\end{equation}

It is easy to verify that for the choice $b_3=-5/12,c_3=-1/24$
 there exists  two global inviscid invariants
\cite{bbkt}:
 the energy, $$E= \sum_{i=1}^{N}(\vert u^+_i \vert^2 +
 \vert u^-_i \vert^2)$$  and helicity, $$H=\sum_{i=1}^{N} k_i
(\vert u^+_i \vert^2 - \vert u^-_i \vert^2).$$

The equations for the fluxes throughout shell number $n$ are:
\begin{eqnarray}
{d \over {dt}}\sum_{i=1,n}E_i&=& k_n 
\left\langle(uuu)_n^E \right\rangle-\nu k_n^2 \sum_{i=1,n}E_i +  E_{in},
\label{flussi0}\\
{d \over {dt}}\sum_{i=1,n}H_i&=& k_n^2 
\left\langle(uuu)_n^H \right\rangle  -\nu  k_n^2\sum_{i=1,n}H_i+ H_{in},
\label{flussi}
\end{eqnarray}
where $E_i$ and $H_i$ are  the  energy and helicity of the $i$-{\it th} shell,
respectively:
 $E_i= \left\langle \left| u_i^+ \right|^2 +\left|u_i^-\right|^2
 \right\rangle$, $H_i =k_i \, \left\langle \left|u_i^+\right|^2 -
\left|u_i^-\right|^2 \right\rangle$.  $E_{in}$ and $H_{in}$ 
are the input of energy and helicity due to forcing effects,
$E_{in}= \Re(\la f^+(u^+_1)^{*}+f^-(u^-_1)^{*}\ra )$, 
$H_{in}=\Re(k_1 \la f^+(u^+_1)^{*}-f^-(u^-_1)^{*}\ra )$.\\ 
In (\ref{flussi0}) and (\ref{flussi})  
we have introduced the triple correlation:
\begin{eqnarray}
\la(uuu)_n^E\ra &=& (\Delta^+_{n+1}+\Delta^-_{n+1})+ 
(b_3+1/2)(\Delta^+_{n}+\Delta^-_{n})
\label{triple0}\\
\la(uuu)_n^H\ra &=& (\Delta^+_{n+1}-\Delta^-_{n+1}) +
 (b_3+1/4) (\Delta^+_{n}-\Delta^-_{n})
\label{triple} 
 \end{eqnarray}
and 
\be
\Delta_n^+= \left\langle \Im(u^-_{n+1}u^+_nu^+_{n-1})\right\rangle.
\label{deltagamma}
\ee
Assuming 
that there exists a stationary state  we have 
${d\over dt} \Pi^{E}_{n}={d \over dt} \Pi^{H}_{n}=0$,
where $\Pi_{n}^{E}=k_{n}\langle (uuu)_{n}^{E}\rangle$ and
$\Pi_{n}^{H}=k_{n}^{2}\langle (uuu)_{n}^{H}\rangle$. Moreover,
in the inertial range we can neglect the viscous contribution 
in (\ref{flussi0}) and (\ref{flussi}), obtaining:
\be
\la(uuu)_n^E\ra=k_n^{-1}E_{in}, \;\;\;
\la(uuu)_n^H\ra=k_n^{-2}H_{in}.
\label{45}
\ee
In \cite{bpt} we have shown that in the case of a large scale forcing 
breaking  inversion 
symmetry (i.e. $f^+ \neq f^-$)  energy and helicity 
fluxes coexist in the systems, both being Reynolds-independent.
In the inertial range we may therefore neglect the viscous contribution
and we obtain for the energy-triple-correlations and for the
helicity-triple-correlation:
\be
\la(uuu)_n^E\ra \sim k_n^{-1},\;\;\;
\la(uuu)_n^H\ra \sim k_n^{-2}.
\label{scaling}
\ee
Relation (\ref{45}) is the equivalent of what found for helical Navier-Stokes 
turbulence in \cite{procaccia2,russo}.

Let us remark that the coexistence of both energy and helicity 
fluxes  is only possible due to the non-positiveness
 of helicity; 
in 2D turbulence, for example,  
a similar result, concerning enstrophy and energy cascades,
 is clearly apriori forbidden. 
 
In \cite{bpt} we have measured the 
 statistics of energy and helicity transfers, by defining 
the two sets of scaling exponents:
\begin{eqnarray}
\Sigma_{E}^{(p)}&\equiv&\la \left((uuu)_n^E \right)^{(p/3)}\ra 
\sim k_n^{-\zeta(p)}\\
\Sigma_{H}^{(p)}&\equiv&\la \left((uuu)_n^H \right)^{(p/3)}\ra
 \sim k_n^{-\psi(p)}.
\label{moments}
\end{eqnarray}

\begin{figure}[htbp]
\centering\epsfig{file=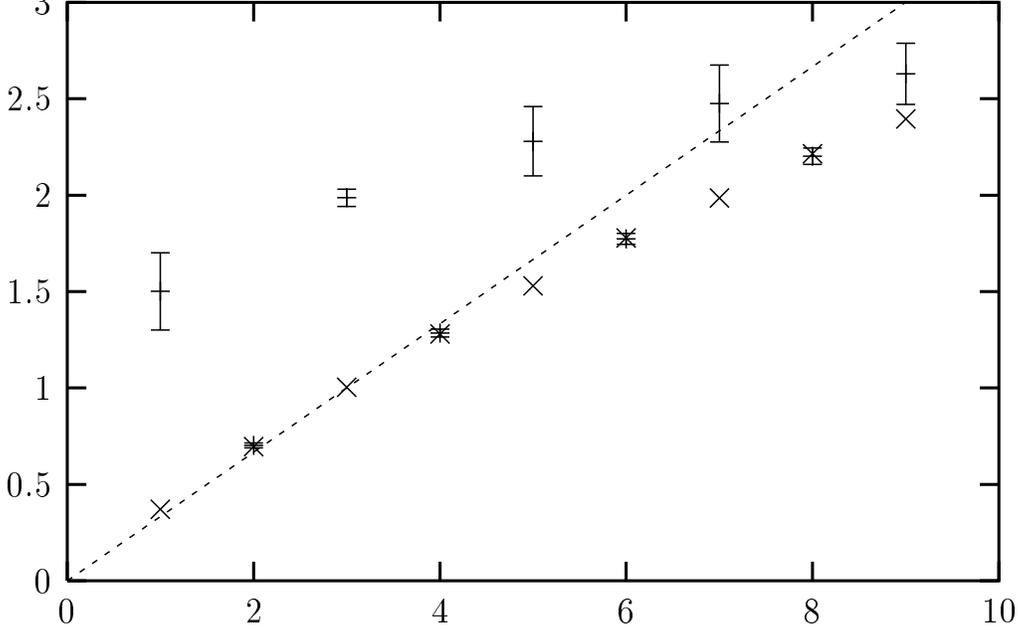,width=0.75\linewidth}
\caption{Anomalous exponents for the helicity flux,
 $\psi_{p}$ ($+$), and for  the energy flux, $\zeta_{p}$ ($\times$),
for $p=1,\dots,9$ obtained with  $N=26$ and $\nu=2\cdot 10^{-9}$.}
\protect\label{fig1}
\end{figure}

As one can see in Figure 1  we found that
the even part of the two statistics coincides, i.e. $\zeta(2p)=\psi(2p)$.
On the other hand, the  scaling exponents of odd moments are different.
In particular the helicity exponents show a strong intermittent 
shape reminding to what one finds, for example, for the statistics
of a passive scalar advected by a turbulent flow.
Phenomenological and numerical evidences supporting the
idea that helicity may be safely be considered as a passive advected
quantity, at small enough scales, have already been 
presented in \cite{orszag}.  

\section{Helicity advection}

Let us push further this idea by analyzing in more details the 
equation of motion describing the evolution of shell variables. 
In order to  highlight the energy and helicity dynamics
it is  better to project the equation of motion into two  new variables
feeling the even and the odd part of the statistics respectively:
\be
w_n =  {{\unp + \unm} \over 2} \,\,\,\, \lambda_n={{\unp - \unm} \over 2}.
\ee
Let us notice that due to the transformation properties with respect
to the symmetry $u_n^+ \leftrightarrow u_n^-$ it is obvious that 
 $w_n$ will mainly feel the energy statistics while $\lambda_n$
will be  highly sensible to the helical properties of the flow. 
The total energy and helicity in the shell $n$ will be
$E_n \propto |w_n|^2 +| \lambda_n|^2$ and $H_n \propto \Re(w_n \lambda_n^*)$.
 \\
Writing now the inertial-time evolution for these two variables we obtain:
\begin{eqnarray}
\label{eq1}
\dot w_n = \left( w_{n+2} w_{n+1} + b_3 w_{n+1}w_{n-1} +c_3 w_{n-1}w_{n-2}
 \right) +\\
\nonumber  \left( - \lambda_{n+2} \lambda_{n+1} -b_3 
\lambda_{n+1}\lambda_{n-1} +c_3\lambda_{n-1}\lambda_{n-2} \right)
\end{eqnarray}
\begin{eqnarray}
\label{eq2}
\dot \lambda_n = \left( w_{n+2} \lambda_{n+1} + b_3 w_{n+1}\lambda_{n-1} -c_3 
w_{n-1}\lambda_{n-2} \right) +\\
\nonumber \left( - \lambda_{n+2} w_{n+1} -b_3 \lambda_{n+1}w_{n-1}
 -c_3\lambda_{n-1}w_{n-2} \right).
\end{eqnarray}
Equations (\ref{eq1}) and (\ref{eq2}) are identical to the set of coupled
equations describing - in the shell world - 
 an active scalar $\lambda_n$ advected by a
turbulent velocity field $w_n$. The active character of $\lambda_n$ is clearly
dictated by the quadratic terms in the RHS of (\ref{eq1}). \\

We now want to argue that at small enough scale the $O(\lambda^2)$
part in the RHS of (\ref{eq1}) becomes negligible and therefore we
end up with a set of eqs describing the passive advection of the
field $\lambda_n$ by the velocity field $w_n$.\\

Let us start  by noticing that for  the variables
$\Delta^{\pm}$ which enter in the definition of the fluxes we have:
\be
\Delta^+ + \Delta^- \propto w_{n+1} \left[ w_n w_{n-1} - \lambda_n 
\lambda_{n-1} \right] + \lambda_{n+1} 
\left[ w_n \lambda_{n-1} - \lambda_n w_{n-1} \right]
\ee
and 
\be
\Delta^+ - \Delta^- \propto  w_{n+1} 
\left[ w_n \lambda_{n-1} - \lambda_n w_{n-1} \right]
 + \lambda_{n+1} \left[ w_n w_{n-1} - \lambda_n \lambda_{n-1} \right].
\ee
Furthermore, by imposing now two general scaling laws for the
field $w_n \sim k_n^{-\alpha}$ and $\lambda_n \sim k_n^{-\beta}$
we obtain:
\be
\left\langle \left( uuu \right)^E_n \right\rangle \sim \Delta^+ + \Delta^- 
\sim w_n^3 + w_n \lambda_n^2 +w_n^2 \lambda_n \sim k_n^{-3\alpha} +k_n^{-\alpha
-2\beta} + k_n^{-2\alpha -\beta}
\label{eq1m}  
\ee
and 
\be
\left\langle \left( uuu\right)^H_n \right\rangle \sim \Delta^+ - \Delta^- 
\sim w_n^2 \lambda_n + w_n\lambda_n^2\sim k_n^{-2\alpha -\beta} +k_n^{-\alpha
-2\beta}.
\label{eq2m}
\ee
By requiring now the asymptotic matching of (\ref{eq1m}) and of
(\ref{eq2m}) with the known behavior for the energy and helicity
fluxes we have:
\be
\left\langle \left( uuu \right)^E_n \right\rangle \sim
 k_n^{-3\alpha} +k_n^{-\alpha -2\beta} + k_n^{-2\alpha -\beta} \sim k_n^{-1}
\label{eq3m}
\ee
\be
\left\langle \left( uuu\right)^H_n \right\rangle \sim
 k_n^{-2\alpha -\beta} +k_n^{-\alpha -2\beta} \sim k_n^{-2}.
\label{eq4m}
\ee
In order to satisfy both matchings  we must  to require that
\be
 w_n \sim k_n^{-1/3} \,\,\, \lambda_n \sim k_n^{-4/3}.
\label{eq5m}
\ee

\begin{figure}[htbp]
\centering\epsfig{file=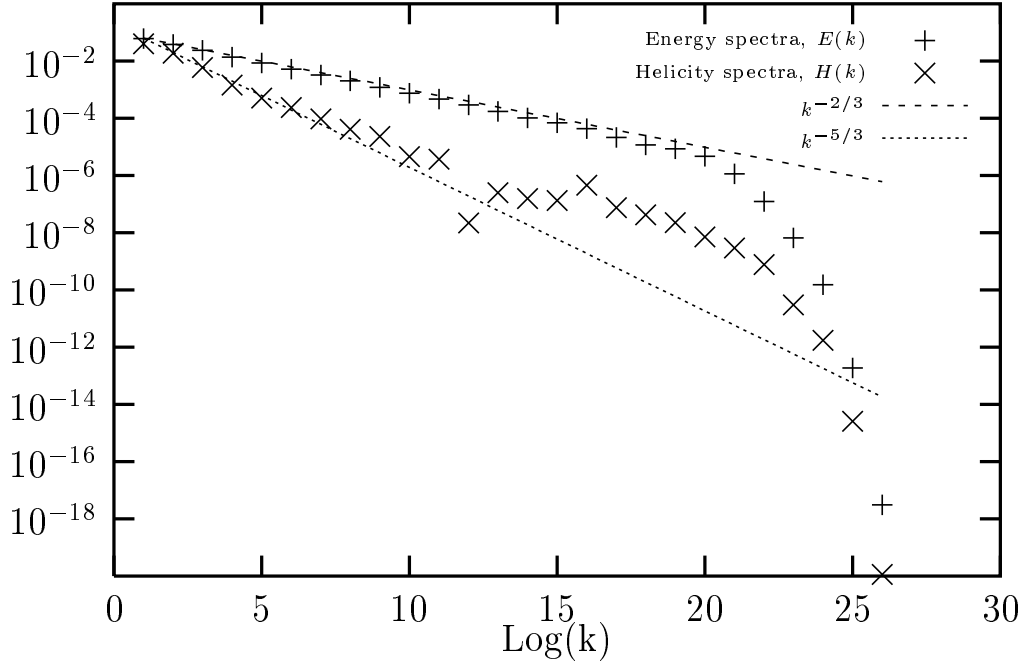,width=0.75\linewidth}
\caption{Log-log plot of Energy, $E(k)$, and helicity, $H(k)$, 
spectra compared with theoretical expectactions.}
\protect\label{fig2}
\end{figure}
By plugging the above scalings in the equation of motion
we discover that  the time evolution
for the field $w_n$ is mainly governed by the field itself, i.e.
the coupling with the field $\lambda_n$ is sub-dominant at small scales
and therefore as a direct consequence 
that the field $\lambda_n$ is passively advected by $w_n$.
Of course the phenomenology here discussed will be 
affected by intermittency, i.e. the above exponents  may show
some weak/strong deviations from the predicted values, without,
however having the possibility to change the  r\^ole
of dominant and sub-dominant contribution in expression 
(\ref{eq3m}) and (\ref{eq4m}).

In order to test this prediction we plot in Figure 2 the energy and helicity
spectrum respectively $E(k_n)/k_n \sim w_n^2 \sim k_n^{-2/3}$ and
$H(k_n)/k_n \sim w_n \lambda_n \sim k_n^{-5/3}$. As one
can see, a part from sthe strong bottleneck effect showed by the helicity
spectrum,  the agreement  is perfect. 

In conclusion we have investigated the 
Helicity transfer statistics in a shell model of turbulence.
In particular, we have found that the strong intermittent
properties of Helicity flux -measured numerically-  can  be explained 
in terms of the  phenomenological evidence that Helicity becomes
 a passive quantity  advected -at
small enough scale- by the energy flux. 


\end{document}